\documentclass{ws-mpla}
\usepackage[super]{cite}
\usepackage{graphicx}
\begin{document}

\markboth{Steven D. Bass}
{Vacuum energy and the cosmological constant puzzle}

%%%%%%%%%%%%%%%%%%%%% Publisher's Area please ignore %%%%%%%%%%%%%%
\catchline{}{}{}{}{}
%%%%%%%%%%%%%%%%%%%%%%%%%%%%%%%%%%%%%%%%%%%%%%%%%%%%%%%%%%%%%%%%%%%

\title{VACUUM ENERGY AND THE COSMOLOGICAL CONSTANT}

%
%INSTRUCTIONS FOR TYPESETTING MANUSCRIPTS\\
%USING \TeX\ OR \LaTeX\footnote{For the title, try not to use %more than
%three lines. Typeset the title in 10 pt Times Roman, 
%uppercase and boldface.}
%}

\author{\footnotesize STEVEN D. BASS
%\footnote{
%Typeset names in 8 pt Times Roman, uppercase. Use the 
%footnote to
%indicate the present or permanent address of the author.}
}

\address{
Stefan Meyer Institute for Subatomic Physics, \\
Austrian Academy of Sciences,
Boltzmanngasse 3, 1090 Vienna, Austria \\
Steven.Bass@assoc.oeaw.ac.at}

\maketitle

\pub{Received (Day Month Year)}{Revised (Day Month Year)}

\begin{abstract}
The accelerating expansion of the Universe points to a small
positive value for the cosmological constant or vacuum energy
density.
We discuss recent ideas that the cosmological constant plus
LHC results might hint at critical phenomena near the Planck scale.
\end{abstract}

\ccode{PACS Nos.: 11.15.Ex, 95.36.+x, 98.80.Es}

\section{Introduction}

Accelerating expansion of the Universe was discovered 
in the observations of distant 
Supernovae 1a \cite{supernova:riess,supernova:perlmutter}, 
and recognised by the 2011 physics Nobel Prize. 
Interpreted within Einstein's theory of General Relativity, the accelerating expansion of the Universe 
is driven by 
a small positive cosmological constant 
or vacuum energy density
perceived by gravitational interactions called dark energy,
for reviews see Refs.~3--18.
Cosmology observations point to an energy budget of 
the Universe where just 5\% is composed of atoms, 26\% involves dark matter
(possibly made of new elementary particles) and 69\% 
is dark energy~\cite{planck15a}.
The vacuum dark energy density extracted 
from astrophysics is 10$^{56}$ 
times smaller than the value expected from the Higgs 
potential in Standard Model particle physics,
which also comes with the opposite negative sign.
Understanding this vacuum energy is an important challenge 
for theory and connects the Universe on cosmological 
scales (the very large) with subatomic physics (the very small).

What might dark energy be telling us about 
the intersection of particle physics and gravitation?
In this paper we focus on the interface of spontaneous
symmetry breaking, vacuum energy and possible critical
phenomena close to the Planck scale (Section 3).
We first briefly review key issues in dark energy science 
and
the vacuum energy in particle physics (Sections 1 and 2).
General Relativity and the Standard Model of particle 
physics work excellently everywhere they have been tested 
in experiments. 
Complementary ideas on 
the cosmological constant are surveyed in Section 4. 
Conclusions are given in Section 5.

The simplest explanation of dark energy is a small positive value for the cosmological constant in Einstein's equations 
of General Relativity.
Einstein's equations link the geometry of spacetime 
to the energy-momentum tensor
\begin{equation}
R_{\mu \nu} - \frac{1}{2} g_{\mu \nu} R
=
- \frac{8 \pi G}{c^2} T_{\mu \nu} + \Lambda g_{\mu \nu} .
\end{equation}
Here $R_{\mu\nu}$ is the Ricci tensor which is built 
from the metric tensor $g_{\mu \nu}$ and its derivatives, 
$R$ is the Ricci scalar and $T_{\mu \nu}$ 
is the energy-momentum tensor.
The left-hand side describes the geometry and 
the right-hand side describes 
the energy content of the gravitational system.
Writing
\begin{equation}
\Lambda = 8 \pi G \rho_{\rm vac} + \Lambda_0 ,
\end{equation}
the cosmological constant tells us about 
the energy density of 
the vacuum $\rho_{\rm vac}$ perceived by gravitational interactions
$\langle T_{\mu\nu} \rangle_{\rm vac}
=
- \frac{g_{\mu \nu}}{c^2} \rho_{\rm vac}$; 
$\Lambda_0$ a possible counterterm.
Being proportional to $g_{\mu \nu}$, a positive cosmological constant corresponds to negative pressure in the vacuum perceived by gravitational interactions.
If the net vacuum energy is finite it will have gravitational effect.
The vacuum energy density receives possible contributions 
from the zero-point energies of quantum fields and 
condensates associated with spontaneous symmetry breaking.

The vacuum is associated with various condensates.
The QCD scale associated with quark and gluon confinement is around 1 GeV 
while the electroweak scale associated with 
the W$^{\pm}$ and Z$^0$ boson masses is around 250 GeV.
These scales are many orders of magnitude less than the 
Planck-mass scale $1.2 \times 10^{19}$ GeV.
If the net vacuum energy is finite it will have gravitational effect.
The vacuum energy density associated with dark energy 
measured in astrophysics experiments
is characterised 
by a scale around 0.002 eV, 
typical of the range of possible 
light neutrino masses, and a cosmological constant
which is 56 orders 
of magnitude less than the value expected from the Higgs condensate with no extra new physics. 
Why is this vacuum ``dark energy" density finite and positive, and why so very small?

There is no strong evidence in the present data 
that dark energy is anything other than a time independent 
cosmological constant.
The most recent Planck measurements of the Cosmic Microwave Background (CMB)
point to a Universe that is spatially flat to an accuracy of 
0.5 \% \cite{planck15a},
consistent with the 
$\Lambda$CDM 6 parameter standard cosmological model.
The value quoted by Planck for the ratio of 
vacuum pressure to dark energy density 
$w = p / \rho$ 
assuming time independent dark energy
is $w = -1.006 \pm 0.045$.
The next generation of experiments
will investigate possible time dependence in the 
dark energy equation of state as well as making 
new precision large distance tests of General Relativity.

The Universe appears to good description as homogeneous and isotropic with
Friedmann-Lemaitre-Robertson-Walker (FLRW)
metric
\begin{equation}
ds^2
=
c^2 dt^2 - a^2(t) \biggl[ \frac{dr^2}{1 - K r^2}
+ r^2 (d \theta^2 + \sin^2 \theta d \phi^2) \biggr] .
\end{equation}
Here 
$a(t)$ is the scale factor which tells us about the sizes of spatial surfaces with $t$ the cosmic time;
$K$ is the three-space curvature constant
($K$ = 0,+1,-1 for a spatially flat, closed or open Universe).
For a flat FLRW Universe
consisting just of matter, radiation or a finite cosmological constant, the energy densities and scale factor behave in an expanding Universe as
\begin{eqnarray}
& & 
\rho_{\rm matter} \sim a^{-3},  
\ \ \ \ a(t) \propto t^{\frac{2}{3}} 
\nonumber \\
& & 
\rho_{\rm radiation} \sim a^{-4}, 
\ \ \ \ a(t) \propto t^{\frac{1}{2}}
\nonumber \\
& & 
\rho_{\rm vac} 
= \frac{\Lambda}{8 \pi G} \sim a^0, 
\ \ \ \ \ a(t) \propto e^{H t}
\end{eqnarray}
where
$
H = {\dot a}/a$ is the Hubble constant.
Condensates that fill all space contribute 
to $\rho_{\rm vac}$. 
Where they are confined within matter 
their contribution forms part of $\rho_{\rm matter}$ 
\cite{brodsky}.
When the density of matter (including both visible and possible dark matter) dominates, the expansion decelerates due to normal gravitational attraction. When the Universe expands to the point that matter becomes dilute and the matter density falls below the vacuum energy density, then the expansion of the Universe changes from deceleration to acceleration.  Supernovae 1a observations tell us that this occurred about five billion years ago, corresponding to redshift about one. 
Given their very different time dependence, it is 
interesting that the matter and dark energy contributions 
to the energy budget of the Universe should be so similar 
at the present time.
Is there something ``special" about ``today"?
Weinberg~\cite{weinberg87}
has argued that (large scale) structure formation
stops when $\rho_{\rm vac}$ starts dominating. 
If $\rho_{\rm vac}$ were too large, there would be no galaxies.

In addition to trying to understand the cosmological constant
puzzle in terms of vacuum energy,
there is also vigorous theoretical activity aimed at
understanding dark energy either by introducing a 
(time dependent) ultra-light
scalar field with finite vacuum expectation 
value to describe the evolution of dark energy in the vacuum 
or, alternatively, modification of long range gravitation
to describe the accelerating expansion of the Universe.
These approaches commonly assume that particle 
physics contributions to the vacuum energy are cancelled 
by some (unknown) symmetry or gravitational counterterm
and then try to interpret the dark energy in terms of the 
new model dynamics.

Each scenario comes with its own theoretical and phenomenological challenges \cite{lesgourgues}.

General Relativity has proved very successful everywhere 
the theory has been tested 
from distances of micrometers through the solar system 
to extra galactic measurements \cite{gravtests}.
At short distances, recent torsion balance experiments
\cite{Kapner:2007}
have found that Newton's Inverse Square Law holds down 
to a length scale of 56 $\mu$m.
Comparable precision is achieved in experiments with 
ultracold neutrons \cite{abele}, 
with the next generation of experiments targeting length
scales in the range 0.1 -- 100 $\mu$m \cite{abele2}.
Precision tests of General Relativity observables in the
strong field regime of double pulsars have been verified 
at the level of 0.05\% \cite{Burgay:2006}.
Studies of gravitational lensing from distant galaxies 
are also in very good agreement with General Relativity predictions \cite{Reyes:2010}.

The expanding Universe might be associated with time 
dependent dark energy 
\cite{solahf15,wetterich14}, 
perhaps connecting the present period of accelerating expansion with initial inflation.
Time dependent dark energy might, in turn, be associated 
with time variation in other 
fundamental constants~\cite{solahf},
{\it e.g.} the fine structure constant $\alpha$,
the ratio of 
electron to proton masses $\mu_{ep}$
(which measures the ratio of the electroweak to QCD scales),
and/or Newton's constant $G$.
For a flat FLRW Universe 
Einstein's equations give~\cite{solahf}
\begin{equation}
\frac{d}{dt} [G (\rho_{\rm matter} + \rho_{\rm vac})]
+
3 \ G \ H \ ( \rho_{\rm matter} + \rho_{\rm vac} ) 
= 0
.
\end{equation}
Experiments give strong constraints 
\cite{calmet,Chiba}
on the possible time dependence of $\alpha$ and $\mu_{ep}$ 
from precision quantum optics experiments (time = today)
\cite{blatt,rosenbland,godun},
molecular clouds in space (time = some billion years ago)
\cite{ubachs13,ubachs15}
and the Cosmic Microwave Background 
(when the Universe was
380 000 years old) \cite{planck}.
Quantum optics measurements give
$
{\dot \alpha} / \alpha = 
(-1.6 \pm 2.3) \times 10^{-17}$~yr$^{-1}$
\ \cite{rosenbland}
and 
the combination
$
{\dot \alpha} / \alpha = 
(-0.7 \pm 2.1) \times 10^{-17}$~yr$^{-1}$
and 
$
{\dot \mu_{ep}} / \mu_{ep} 
 = (0.2 \pm 1.1) \times 10^{-16}$~yr$^{-1}$
\ \cite{godun},
where the latter measurement 
assumes time constancy of nuclear magnetic moments.
From molecular clouds in space 
(time = 7.5 billion years ago) 
${\dot \mu_{ep}} / \mu_{ep} < 2 \times 10^{-17}$~yr$^{-1}$
\ \cite{ubachs13}.
Planck measurements of the CMB~\cite{planck}
give $\alpha / \alpha_0 = 0.9989 \pm 0.037$.
The most sensitive parameter to possible change in the dark energy density is $\mu_{ep}$. 
Time dependence of Newton's constant is constrained 
from a range of experiments from the solar system 
to the CMB with measured
bound typically about ${\dot G}/G < 10^{-11}$~yr$^{-1}$
\cite{Chiba}. 
Fritzsch and Sola~\cite{solahf}
emphasise that
the present experimental bounds on change of the nucleon mass and $\Lambda_{\rm QCD}$ are compatible with the corresponding bounds on cosmic evolution of $G$ and $\rho_{\rm vac}$.
Time dependent couplings are commonly interpreted 
in the theoretical literature in terms of the time 
dependence of some new scalar field which couples to matter 
\cite{barrow}.

If we model dark energy by the matrix element of a time dependent scalar field, one requires that this scalar has 
very small mass, 
today of order $10^{-33}$ eV, with Compton wavelength
bigger than the inverse Hubble radius to avoid 
clumping and to ensure 
uniform distribution in the present Universe
\cite{peebles88,wetterich88}.
What protects this 
tiny mass from quantum radiative corrections? 
Coupling a near massless scalar to Standard Model particles will introduce a ``fifth force" (which is not gauged unlike the other forces of nature). 
At the present time there is no experimental evidence 
for any such interaction so couplings between 
the new scalar and matter must be very much suppressed.
Wetterich et al.~\cite{wetterich14a}
have argued that dark energy to neutrino coupling
might lead to neutrino lumps 
that might be looked for in future astrophysics experiments.
Coupling to a time dependent scalar field will
in general induce time dependence in the fundamental constants.

\section{Vacuum energy and the cosmological constant}

The vacuum energy receives possible contributions 
from
the zero-point energy associated with quantisation
as well as
condensate contributions 
induced by the Higgs mechanism and dynamical symmetry breaking.

Quantisation introduces zero-point vacuum energies 
for quantum fields and therefore, in principle, 
can affect the geometry through Einstein's equations.
Before normal ordering the zero-point energy of the vacuum is badly divergent, being the sum of zero-point energies for an infinite number of oscillators, one for each normal mode, or 
degree of freedom of the quantum fields \cite{bjd}.
Before interactions, the vacuum (or zero-point) energy is
\begin{equation}
\rho_{\rm vac} =
\frac{1}{2} \sum \{\hbar\omega \}
=
\frac{1}{2} \hbar
\sum_{\rm particles} g_i \int_0^{k_{\rm max}}
\frac{d^3 k}{(2 \pi)^3} \sqrt{k^2 + m^2}
.
\end{equation}
Here $\frac{1}{2} \{ \hbar \omega \}$ 
denotes the eigenvalues of the free Hamiltonian and
$\omega = \sqrt{k^2 + m^2}$
where $k$ is the wavenumber and $m$ is the particle mass;
$g_i = (-1)^{2j} (2j+1)$
is the degeneracy factor for a particle $i$ of spin $j$, 
with $g_i  >0$ for bosons and $g_i < 0$ for fermions.
The minus sign follows from the Pauli exclusion principle and 
the anti-commutator relations for fermions.

The mass scale appearing in the zero-point energy depends 
on the ultraviolet regularisation.
Eq.(6) for $\rho_{\rm vac}$ 
is quartically divergent in the cut-off $k_{\rm max}$. 
If we take $k_{\rm max}$ of order the Planck scale
where we expect quantum gravity effects to become important,
$M_{\rm Pl} 
= \sqrt{ \hbar c / G } = 1.2 \times 10^{19} \ {\rm GeV}$,
then we obtain a value 
for $\rho_{\rm vac}$ which is $10^{120}$ times too big.
Zero-point energies are, in themselves, not Lorentz covariant without a corresponding vacuum pressure, 
$\rho_{\rm vac} = - p_{\rm vac}$. 
If one instead evaluates the integral in 
Eq.(6) using dimensional regularisation 
$\overline{\rm MS}$ instead of a cut-off on the 3-momentum, 
then one finds~\cite{jmartin}
\begin{equation}
\rho_{\rm vac} = - p_{\rm vac}
\simeq 
- 
\frac{1}{2} \hbar \ g_i \
\frac{m^4}{64 \pi^2} 
\biggl[ \frac{2}{\epsilon} + \frac{3}{2} - \gamma
- \ln \biggl( \frac{m^2}{4 \pi \mu^2} \biggr) \biggr]
+ ...
\end{equation}
for the contribution from particles with mass $m$,
that is 
proportional to the fourth power of the particle 
mass instead of the ultraviolet cut-off $k_{\rm max}$.
Zero-point contributions would cancel in a world with exact supersymmetry because of the sign change between boson and
fermion contributions in Eq.(6).

In quantum field theory (without coupling to gravity) 
the zero-point energy is removed by normal ordering 
so that the zero of energy is defined as the energy of 
the vacuum.
This can be done because absolute energies here are not measurable observables. 
Before we couple to gravity, only energy differences have physical meaning, 
{\it e.g.} in Casimir processes ~\cite{astrid,jaffe} 
which measure the force between parallel conducting plates 
in QED and which contribute a ``cavity term'' to the mass 
of the proton in Bag models of quark confinement.
The net vacuum energy is measured only through large distance gravity and astrophysics.

Suppose one can argue away zero-point contributions 
to the vacuum energy.
For example, the Casimir force can also be calculated 
without 
reference to zero-point energies
\cite{jaffe}.
One still has condensates associated with spontaneous 
symmetry breaking.
Condensates which carry energy enter at various energy scales 
in the Standard Model.
The Higgs condensate gives 
\begin{equation}
\rho_{\rm vac}^{\rm ew} \sim - (250 \ {\rm GeV})^4 
\ ,
\end{equation}
56 orders of magnitude larger and with opposite negative
sign to the observed value
\begin{equation}
\rho_{\rm vac} \sim + (0.002 \ {\rm eV})^4 .
\end{equation}
The QCD quark condensate gives about $- (200 \ {\rm MeV})^4$.
If there is a potential in the vacuum it will, 
in general, correspond to some finite vacuum energy.
Why should the sum of many big numbers (plus any possible gravitational counterterm) add up to a very small number?

The Higgs and QCD condensates form at different times 
in the early Universe, suggesting some time dependence 
to $\rho_{\rm vac}$.
Further large condensate contributions might be expected 
also in Grand Unified Theories.

\section{Vacuum energy and high-scale phenomena}

What might the cosmological observations and particle physics being telling us ?

It is interesting that the dark energy or cosmological constant scale $0.002 \ {\rm eV}$
in Eq.(9) is of the same order that we expect 
for the light neutrino mass
~\cite{bass:acta,altarelli,nucond:caldi,fardon04,wetterich07}.
Light neutrino mass values $\sim$ 0.004-0.007 eV 
are extracted from studies of neutrino 
oscillation data assuming normal hierarchy 
and values less than about 0.02 eV are obtained with 
inverted hierarchy \cite{neutrinomass:alt,neutrinomass:hf}.
That is, 
one finds the phenomenological relation
\begin{equation}
\mu_{\rm vac} \sim m_{\nu} \sim \Lambda_{\rm ew}^2/M 
\end{equation}
where $M \sim 3 \times 10^{16}$ GeV 
is logarithmically close to the Planck mass $M_{\rm Pl}$
and
typical of the scale that appears in Grand Unified Theories.
There are also theoretical hints that this large mass scale 
might perhaps be associated with dynamical symmetry breaking,
see below. 
The gauge bosons in the Standard Model which 
have a mass through the Higgs mechanism are also the gauge bosons 
which couple to the neutrino. 
Is this a clue?
The non-perturbative structure of chiral gauge theories is not well understood.
If taken literally Eq.(10) 
connects neutrino physics, 
Higgs phenomenon in
electroweak symmetry breaking and dark energy 
to a new high mass scale which needs to be understood.

We next argue how this physics might be connected, 
first treating neutrino chirality by analogy with 
the ``spins'' in an Ising-like system that becomes 
active near the Planck mass
\cite{bass:acta}.
Then, in Section 3.2 we discuss recent LHC results 
which, 
when evolved to large scales, 
suggest stability or metastability of 
the Standard Model vacuum 
and which might hint at possible critical phenomenon at 
some very large scale.

\subsection{Neutrinos and the subatomic vacuum}

Changing the external parameters of the theory can change 
the phase of the ground state.
For example, QED in 3+1 dimensions with exactly massless electrons is believed to dynamically generate a photon mass \cite{gribov}.
In the Schwinger Model for 1+1 dimensional QED on a circle, setting the electron mass to zero shifts the theory from a confining to a Higgs phase \cite{gross}.

In Standard Model particle physics
QED is manifest in the Coulomb phase,
QCD is manifest in the confinement phase and
the weak interaction is manifest in the Higgs phase.
The W$^{\pm}$ and Z$^0$ gauge bosons 
which have a mass through the Higgs mechanism 
are also the gauge bosons which couple to the neutrino
and the QED photon and QCD gluons are massless. 
What happens to the structure of non-perturbative propagators and vacuum energies when we turn off the coupling of the gauge bosons to left- or right-handed fermions?

Consider Yang-Mills SU(2) with and without parity violation.
Pure Yang-Mills theory and Yang-Mills theory coupled to fermions are  both confining theories but the mechanism is different for each. 
Confinement is intimately connected with dynamical chiral symmetry breaking \cite{thooft}.
Scalar confinement implies dynamical chiral symmetry breaking and a fermion condensate 
$\langle {\bar \psi} \psi \rangle < 0$.
This scalar condensate is absent if there is no right-handed fermion
participating in the interaction.
Suppose that the theory is ultraviolet consistent,
{\it e.g.} that it is embedded in a larger theory 
to ensure anomaly cancellation necessary for gauge 
invariance and
renormalisability.
Switching off the coupling of SU(2) gauge bosons 
to right-handed fermions must induce some modification of 
the non-perturbative propagators.
Either confinement is radically reorganised or
one goes to a Coulomb phase or to a Higgs phase
whereby the Coulomb force is replaced by a force of finite range with finite mass scale and the issues associated with infrared slavery are avoided.

We next suppose the confinement to Higgs transition applies.
That is, we suppose that the non-perturbative 
ground state of chiral gauge theory is in a Higgs phase.
Anomaly cancellation in the ultraviolet 
is required by gauge invariance and renormalisability.
If some dynamical process acts to switch off 
the coupling of left- or right-handed fermions,
it will
have important consequences for the theory 
in the ultraviolet limit and should therefore be active there.
If symmetry breaking is dynamical and hence non-perturbative 
it will appear with coefficients smaller than any power of the running coupling. 
Suppose an exponentially small effect \cite{witten}.
Dynamical symmetry breaking then naturally induces a 
symmetry breaking scale $\Lambda_{\rm ew}$ 
which is much smaller than the high energy scales in the problem 
$M_{\rm cutoff}$.
If we take the mass scale $M_{\rm cutoff}$ 
to be very large, {\it e.g.} close to the Planck scale,
then the expression
\begin{equation}
\Lambda_{\rm ew}
 = M_{\rm cutoff} \ e^{-c/g(M_{\rm cutoff}^2)^2} 
\ \ll \ M_{\rm cutoff}
\end{equation}
naturally leads to hierarchies.
Symmetry breaking effects at very large scales are suppressed 
by the exponential with the result that
$\Lambda_{\rm ew}$ is the mass scale appearing in the particle theory which describes the energy domain probed in laboratory experiments.
In Eq. (11) if we take the ratio of the weak scale 
$\Lambda_{\rm ew}$ to the mass scale in Eq.(10), then
$\Lambda_{\rm ew}/M \sim 10^{-14}$.
If we take the ultraviolet mass scale to be the Planck mass,
then
$\Lambda_{\rm ew}/M_{\rm Pl} \sim 10^{-17}$.

We next consider a phenomenological trick to investigate 
the different scales in the problem.
Analogies between quantum field theories and condensed matter and statistical systems have often played an important role in 
motivating ideas in particle physics.
Here we consider a possible analogy between the neutrino vacuum and the Ising model of statistical mechanics where 
the ``spins'' in the Ising model are associated with neutrino chiralities \cite{bass:acta}. 
The free energy for the statistical ``spin'' system plays the role of the vacuum energy density in quantum field theory \cite{kogut}.

The ground state of the Ising model exhibits spontaneous magnetisation where all the spins line up and
the internal 
energy per spin and the free energy density of the spin 
system go to zero
with corrections dampened by the exponential factor 
$e^{- \beta J}$.
Here $J$ is the spin-spin 
coupling in the Ising Hamiltonian
\begin{equation}
H = 
- J \ \sum_{i,j} \ 
( \sigma_{i,j} \sigma_{i+1,j} + \sigma_{i,j+1} \sigma_{i,j} )
\  ,
\end{equation}
$\beta = \frac{1}{k T}$ where
$k$ is Boltzmann's constant and $T$ is the temperature. 
For an Ising system with no external magnetic field
the free energy density is equal to minus the pressure
\begin{equation}
P = - \biggl( \frac{\partial F}{\partial V} \biggr)_{T}
\end{equation}
-- that is, the model equation of state looks like a vacuum energy term in Einstein's equations of General Relativity, $\propto g_{\mu \nu}$.

We take $J \sim + M$ to be large and close to the Planck mass. 
The exponential suppression factor $e^{-2 \beta J}$ then
ensures that non-renormalisable fluctuations associated 
with the Ising-like interaction are negligible in the 
ground state, 
which is in the spontaneous magnetisation phase involving 
just left-handed ``neutrinos''.
Following our previous discussion, it seems reasonable 
to believe that the SU(2) gauge symmetry coupled 
to the neutrino is now spontaneously broken, 
that is the SU(2) gauge symmetry associated with 
the W$^{\pm}$ and Z$^0$ bosons is in the Higgs phase.

Weak interactions mean that we have two basic scales in the problem: $J \sim M$ and the electroweak scale 
$\Lambda_{\rm ew}$ induced by spontaneous symmetry breaking.
For a spin model type interaction, the ground state with left-handed ``spin'' chiralities is characterised by vanishing energy density. Excitation of right-handed chiralities is associated with the large scale $2M$.
Then the mass scale associated with the vacuum for the ground state of the combined system (spin model plus gauge sector) 
one might couple to gravity reads in matrix form as 
\begin{equation} 
\mu_{\rm vac}
\sim 
\left[ \begin{array}{cc}
\! 0  &  - \Lambda_{\rm ew}  \! \\
\! - \Lambda_{\rm ew} &  -2M  \!
\end{array} \right] 
\end{equation}
with the different terms depending how deep we probe into the Dirac sea.
Here the first row and first column refer to left-handed states of the spin model ``neutrino'' and the second row and second column refer to the right-handed states. 
The off-diagonal entries correspond to the potential in the 
vacuum associated with the dynamically generated Higgs sector.
Eq.(14) looks like the see-saw mechanism 
\cite
{seesaw:mink,seesaw:gellmann,seesaw:yanagida,seesaw:mohapatra} 
proposed to explain neutrino masses. 
Diagonalising the matrix for $M \gg \Lambda_{\rm ew}$ 
gives the light mass eigenvalue 
\begin{equation}
\mu_{\rm vac} 
\sim  \Lambda^2_{\rm ew} / 2M
\end{equation}
-- that is, the phenomenological result in Eq.(10).
Here the electroweak contribution $\Lambda_{\rm ew}$ 
is diluted by the ``spin'' potential in the vacuum.
The resultant picture~\cite{bass:acta}
is a Higgs sector characterised 
by scale $\Lambda_{\rm ew}$ embedded 
in the ``spin" polarised ground state that holds up to the ultraviolet scale $2M$.
That is, the Standard Model including QCD acts like an 
``impurity''
in the ``spin'' polarised vacuum.

\subsection{Stability of the Standard Model vacuum}

Results from the LHC experiments ATLAS, CMS and LHCb 
are in good agreement with the Standard Model with 
(so far) no evidence of new physics.
The Higgs boson discovered at 
LHC \cite{higgs:atlas,higgs:cms} is consistent 
with Standard Model expectations \cite{guido}.
It is an open question whether at a deeper level this boson 
is elementary or of dynamical origin. 
Recent precision measurements of the electron 
electric dipole moment, EDM, are consistent with zero
(with upper bound $| d_e | < 8.7 \times 10^{-29}$ $e$ cm), 
constraining possible new sources of CP violation from beyond the Standard Model up to scales similar to or larger than those probed at the LHC \cite{hinds,acme}.
The next generation electron EDM experiments expect to probe up
to the 100 TeV scale.
Precision measurements of the neutron and nuclear EDMs 
\cite{musolf}
and tests of CPT and Lorentz invariance 
\cite{kostelecky}
are so far all consistent with the Standard Model.

It is interesting to consider the possibility that 
the Standard Model might work up to a scale close 
to the Planck mass \cite{shaposhnikov}
with stable or metastable vacuum, 
as well
the implications for possible deeper structure 
and the 
vacuum energy puzzle. 
Possibly, the Standard Model might be emergent as the 
long range tail of some critical Planck system \cite{fredj1}.
Whilst the Standard Model has proved very successful everywhere it has been tested 
we know that some extra physics needed to explain 
the very small neutrino masses, the baryon asymmetry and
strong CP problems as well as dark matter and inflation.
The scale of this new physics is as yet unknown
and not yet given by experiments.

Recent perturbative renormalisation group (RG) 
calculations suggest that the Standard Model 
vacuum with the measured Higgs and top quark masses
$m_H = 125.15 \pm 0.24$ GeV
and
$m_t = 173.34 \pm 0.76$ GeV \ \cite{ellis}
might be stable or metastable
(with half-life much greater than 
 the present age of the Universe)
\cite{fredj1,masina,degrassi,buttazzo,bezrukov}.
An unstable vacuum would require some new interaction at
higher scales.
Which scenario occurs is sensitive to technical details 
in calculating $\overline{\rm MS}$ parameters in terms of physical ones and how one should include tadpole diagrams to be consistent with gauge invariance.
The important issue here is that the $\beta$ function 
for the Higgs four-boson self-coupling $\lambda$ has 
a zero and when (if at all) this coupling $\lambda$ 
crosses zero 
(either around $10^9$ GeV \ \cite{degrassi,buttazzo}
or perhaps not at all \cite{fredj1,masina}). 
These calculations assume perturbative evolution of the Standard Model up to the highest possible scales of order 
the Planck mass without coupling to additional 
``new physics'' 
(including any possible quantum gravity 
 or possible dark matter candidates) in the evolution. 
The RG calculations involve three families of fermions active in the RG equations and the Higgs boson is taken as elementary in these calculations.

Vacuum stability is very sensitive to the exact values of
the Higgs and top-quark masses. 
For the measured value of $m_t$, 
$m_H$ is very close to the smallest value to give a stable vacuum with 
the vacuum being at the border of stable and metastable.
With modest changes in $m_t$ and $m_H$
(increased top mass 
 and/or reduced Higgs mass)
the Standard Model vacuum would be unstable 
\cite{fredj1,masina,degrassi,buttazzo,bezrukov}.
If the vacuum is indeed stable up to the Planck mass, 
perhaps there is some new critical phenomena to be understood in the extreme ultraviolet?
One is led to consider the possibility that there is 
no new scale between the electroweak scale and some 
very high scale close to the Planck mass.

Radiative corrections to the Higgs mass in the ultraviolet
are very interesting.
The running Higgs mass $m_H$ is related to the bare mass 
$m_{0 H}$ through
\begin{equation}
m_H^2 = m_{0 H}^2 - \delta m_H^2, 
\ \ \ \
\delta m_H^2 = \frac{M_{\rm Pl}^2}{16 \pi^2} C_1
\end{equation}
where
%\begin{equation}
$
\delta m_H^2
$
is the mass counterterm and
\begin{equation}
C_1 
= \frac{6}{v^2} (M_H^2 + M_Z^2 + 2 M_W^2 - 4 M_t^2)
= 2 \lambda + \frac{3}{2} g'^2 + \frac{9}{2} g^2 - 12 y_t^2
.
\end{equation}
Here $v$ is the Higgs vacuum expectation value, $\lambda$ 
is the Higgs self-interaction coupling, $g'$ and $g$ are 
the electroweak couplings and $y_t$ is the top quark Yukawa coupling.
The small value of 
$m_H^2$ relative to $M_{\rm Pl}^2$ is the hierachy problem
and connected to discussions~\cite{giudice} of naturalness.
Taking the couplings in the formula for $C_1$ to be RG
scale dependent
and the measured Higgs and top quark masses, 
Jegerlehner \cite{fredj1}
has argued that $C_1$ crosses zero at a scale 
$\sim 10^{16}$ GeV, logarithmically close to the Planck mass. 
He argues that 
the sign change in the Higgs bare mass squared
triggers the Higgs mechanism
with a first order phase transition
if the Standard Model is understood as the 
low energy effective theory of some cutoff system residing at
the Planck mass 
\cite{fredj1,fredjhelv,fredj98}.
In this scenario the Higgs might act as 
the inflaton at higher mass scales in a 
symmetric phase characterised 
by a very large bare mass term \cite{fredj2,fredj15}. 
Note that the Higgs and top quark masses 
are taken to be time independent in these calculations.
Further, the electroweak Higgs contribution 
to the vacuum energy density $- \frac{\lambda}{24} v^4$
obeys a similar expression to 
Eq.(17) and crosses zero at a
similar scale about $10^{16}$ GeV 
so the renormalised version of this quantity can be much 
less than the bare version at scales close to the Planck mass.

It is interesting that the scale $\sim 10^{16}$ GeV
found in this calculation also arises 
(modulo Yukawa couplings) 
in the see-saw mechanism for neutrino masses
and 
in the ``spin'' model argument 
for dynamical symmetry breaking in Section 3.1,
as well as in Grand Unified Theories. 
The scale of inflation is related to the tensor to scalar ratio $r$ in B modes in the cosmic microwave background 
through
$
V_{\rm inflation} \sim 
\bigl( \frac{r}{0.01} \bigr)^{\frac{1}{4}} 10^{16} \ {\rm GeV}
$
\ \cite{lyth}.
A finite value of $r$ would be evidence of gravitational 
waves from the inflationary period.
If ongoing and future measurements converge on a positive signal in 
the region $0.001 < r < 0.1$,
then this would point to a scale of 
inflation in the same region close to $10^{16}$ GeV.

The issue of vacuum stability is important. 
If some critical process is at work,
then one might speculate 
that the Standard Model is itself emergent,  
as the long range tail of a critical Planck system.
The emergence scenario differs from the paradigm 
of unification with maximum symmetry at 
the highest possible energies, with a unification big 
gauge group
spontaneously broken through various Higgs condensates
to the Standard Model, with each new condensate 
introducing 
an extra large contribution
to the vacuum energy and the cosmological constant.
In the emergence scenario one might expect violations of
gauge and possibly Lorentz invariance as well as
renormalisability at scales close to the Planck mass.
Perhaps the gauge theories of particle physics and 
also
General Relativity are effective theories with
characteristic energy of order the Planck scale
\cite{weinberg09}.
The idea that local gauge symmetries might be emergent 
dates to early work of Bjorken \cite{bj1963,bj2001}
who suggested that the photon might be a Goldstone boson associated with 
spontaneous breaking of Lorentz invariance.
There are strong experimental constraints on 
possible Lorentz invariance violation~\cite{kostelecky}.
Bjorken has further suggested that any breakdown of
gauge symmetries, 
with the activation of gauge degrees of freedom and a
preferred choice of gauge
associated with emergent gauge symmetry,
might vanish in the limit of vanishing dark energy 
\cite{bj2001,bj2010}.
Emergence ideas are further discussed in 
Refs.~\refcite{fredjhelv,fredj98,hooft07,nielsen,wen}.
Patterns in fermion masses have been interpreted 
\cite{hf79,hf14}
to suggest that perhaps there is a deeper structure 
to matter 
and that perhaps the fermions and W$^{\pm}$ and Z$^0$ 
bosons might be composite.
Perhaps it is possible to re-interpret these ideas 
also in 
terms of an emergent Standard Model?

Ideas about the cosmological constant based on emergence phenomena in condensed matter physics have also been 
suggested~\cite{volovik}.
If the vacuum of particle physics acts like a cold quantum liquid in equilibrium, then its pressure vanishes unless 
it is a droplet in which case there will be surface corrections scaling as an inverse power of the droplet size.
Vacuum dark pressure scales with the vacuum dark energy density and is measured by the cosmological constant which
scales as the inverse square of the Hubble length $R = 1/H$
(or ``size" of the Universe), 
{\it viz.}
$\Lambda 
 = 8 \pi G \rho_{\rm vac} = 3 H^2 = 3/R^2$
in a Universe dominated by dark energy.

\section{Complementary ideas}

We briefly mention other ideas involving the cosmological constant and gravitational dynamics or where $M_{\rm Pl}$
plays a vital role.

Brandenberger et al.~\cite{brandenberger}
and Polyakov \cite{polyakov} 
have argued that 
de Sitter space is unstable in the presence of quantum
fields.
Gravitational waves propagating in a background 
spacetime affect the dynamics of the background. 
Gravitational backreaction might generate a negative contribution to the cosmological constant in the
terminating of
inflation and thus screen the cosmological constant today.

Ward~\cite{ward13,ward14}
considers the cosmological constant in a 
model of resummed quantum gravity with an asymptotically safe
ultraviolet fixed point \cite{weinbergas}.
He finds a value of the cosmological constant 
close to the measured value with theoretical error of 
a factor of $10^4$.

In the causal dynamical triangulation approach 
to quantum gravity Ambjorn et al.~\cite{loll}
start with 
the gravitational path integral 
\begin{equation}
Z (G, \Lambda) = \int {\cal D} g \ e^{i S_{G, \Lambda} [g]}
\end{equation}
before coupling to matter and
taking as inputs causality and locality plus Newton's
constant and the cosmological constant as parameters.
Curved space-time at early intermediate stages in the time
evolution is approximated by triangulations.
This approach generates de Sitter space with an emergent 4
dimensions of space-time (starting from 2 dimensions near
the Planck mass).

In a different approach where the Planck mass also 
plays a vital role,
McLerran et al~\cite{mclerran12}
assume that the sum of baryon and lepton
number might not be conserved at a very high scale near 
the Planck mass through electroweak axion coupling
to the topological charge of the electroweak gauge theory
and instantons. The electroweak axion might then generate 
a dark energy contribution close to the measured value if
there is no new physics between the electroweak and Planck
scales.

\section{Conclusions}

The cosmological constant puzzle continues to fascinate.
Why is it finite, positive and so very small?
What suppresses the very large vacuum energy 
contributions expected from particle physics?
Is the accelerating expansion of the Universe really driven 
by a time independent cosmological constant or by new possibly time dependent dynamics?
Experiments will push the high-energy and precision 
frontiers of subatomic particle physics. 
Is new physics ``around the corner'' or might the
Standard Model work up to a very large scale, 
perhaps close to the Planck mass and perhaps hinting at critical new phenomena in the ultraviolet?
Understanding the accelerating expansion of the Universe and the cosmological constant vacuum energy puzzle promises to teach us a great deal about the intersection of subatomic physics and dynamical symmetry breaking on the one hand, and gravitation on the other.

\section*{\bf Acknowledgements}

I thank S. J. Brodsky, F. Jegerlehner and J. Wosiek for helpful discussions.

The research of SDB is supported by the Austrian Science Fund, 
FWF, through grant P23753.

%%%\newpage

\end{document}